\documentclass[aps,prl,twocolumn]{revtex4-1}

\usepackage{amsmath}
\usepackage{amssymb}
\usepackage{graphicx}
\usepackage{color}
\usepackage[colorlinks,citecolor=blue,linkcolor=red,urlcolor=blue]{hyperref}

\newcommand{\bra}[1]{\langle #1|}
\newcommand{\dd}{\mathrm{d}}

\newcommand{\ket}[1]{|#1\rangle}
\newcommand{\mbb}[1]{\mathbb{#1}}
\newcommand{\mbf}[1]{\mathbf{#1}}

\newcommand{\mr}[1]{\mathrm{#1}}

\begin{document}
	
\title{Low-Temperature Transport Property of Spin-1/2 Random Heisenberg Chains}

\author{Yuejiu Zhao}
\affiliation{Kavli Institute for Theoretical Sciences and CAS Center for Excellence in Topological Quantum Computation, University of Chinese Academy of Sciences, Beijing 100190, China}

\author{Long Zhang}
\email{longzhang@ucas.ac.cn}
\affiliation{Kavli Institute for Theoretical Sciences and CAS Center for Excellence in Topological Quantum Computation, University of Chinese Academy of Sciences, Beijing 100190, China}

\date{\today}

\begin{abstract}
Quenched disorders can strongly influence the physical properties of quantum many-body systems. The real-space strong-disorder renormalization group (SDRG) analysis has shown that the spin-1/2 random Heisenberg chain is controlled by the infinite-randomness fixed point (IRFP) and forms a random singlet (RS) ground state. Motivated by recent thermal transport experiments on the quasi-one-dimensional antiferromagnet copper benzoate [B. Y. Pan et al, Phys. Rev. Lett. {\bf 129}, 167201 (2022)], we adapt the SDRG to study the low-temperature properties of the random Heisenberg chain by assuming that its low-energy excited states are captured by the parent Hamiltonian of the RS ground state as well. We find that while the specific heat coefficient and the uniform magnetic susceptibility scale as $C/T\sim T^{-\alpha_{c}}$ and $\chi\sim T^{-\alpha_{s}}$ with $0<\alpha_{c,s}<1$, indicating a divergent low-energy density of states, the thermal and the spin conductivities scale as $\kappa/T\sim T$ and $\sigma_{s}\sim T$, which implies a vanishing density of extended states in the low-energy limit. We believe that such a disparity in the thermodynamic and transport properties is a common feature of random systems controlled by the IRFP.
\end{abstract}	
	
\maketitle

{\bf Introduction.} Quenched disorders are ubiquitous and result into a plethora of exotic states of matter in quantum materials. A prominent example is the Anderson localization in noninteracting electron systems \cite{Anderson1958, Abrahams1979, Abrahams201050}. In the past decade, intense research efforts have been devoted to the possible many-body localization (MBL) in strongly disordered quantum many-body systems \cite{Nandkishore2015, Abanin2019a}, focusing on highly excited states with finite energy density, though the general conclusion remains debated \cite{Suntajs2020a, Suntajs2020, Sels2023}.

Quenched disorders are also crucial in shaping the ground state and low-energy properties of quantum many-body systems. The real-space strong-disorder renormalization group (SDRG) analysis of the one-dimensional (1D) spin-1/2 random antiferromagnetic (AF) Heisenberg model shows that the disorder distribution becomes broader and broader with the RG transformations, thus the ground state is captured by the infinite-randomness fixed point (IRFP) and approximately given by the direct product of decoupled spin singlet pairs \cite{Ma1979, Dasgupta1980, Fisher1994}. The specific heat coefficient $C/T$ and the uniform magnetic susceptibility $\chi$ diverge in the low-temperature limit irrespective of the initial distribution of exchange interaction strengths \cite{Ma1979, Dasgupta1980, Hirsch1980, Hirsch1980a, Bhatt1981a, Bhatt1982}. The SDRG analysis has been generalized to a large class of random spin models \cite{Igloi2005, Igloi2018a}. Moreover, it is also applied to the dynamical spin correlation and transport properties of random spin chains at zero temperature \cite{Damle2000, Motrunich2001} as well as highly excited states of strongly disordered spin chains in the postulated MBL regime \cite{You2016a, Slagle2016}.

In a recent work \cite{Pan2022}, the low-temperature specific heat $C$ and the thermal conductivity $\kappa$ of the quasi-1D antiferromagnet copper benzoate [Cu(C$_{6}$H$_{5}$COO)$_{2}\cdot$3H$_{2}$O] were studied. This material is composed of weakly coupled chains of spin-1/2 Cu$^{2+}$ ions, and can be described by the AF Heisenberg model \cite{Date1970, Dender1996, Dender1997}. The magnetic contributions to $C$ and $\kappa$ are linearly proportional to the temperature in the intermediate temperature regime, indicating highly conductive spinon excitations. However, $\kappa/T$ drops dramatically at low temperature, while $C/T$ remains almost constant. Such a disparity in the thermodynamic and transport properties was attributed to the possible ``spinon localization'' or MBL induced by unavoidable weak disorders \cite{Pan2022}.

Motivated by the above experimental results, we theoretically study the finite-temperature transport property of the random Heisenberg chain. Assuming that the low-energy excited states can also be approximately captured by the parent Hamiltonian of the ground state obtained with the SDRG transformations, we evaluate the physical quantities in the low-temperature regime. We find significantly different scaling behavior of thermodynamic quantities and transport coefficients. While the specific heat coefficient and the uniform magnetic susceptibility scale as $C/T\sim T^{-\alpha_{c}}$ and $\chi\sim T^{-\alpha_{s}}$ with $0<\alpha_{c,s}<1$, which is consistent with previous SDRG results and the thermodynamic experiments in disordered spin chain materials \cite{Bulaevskii1972, Azevedo1977} and indicates a divergent low-energy density of states, the spin conductivity $\sigma_{s}$ and the thermal conductivity coefficient $\kappa/T$ diminish in the low-temperature limit, $\sigma_{s}\sim T$ and $\kappa/T\sim T$. Such a disparity implies that most of the low-energy excited states are essentially localized. We believe this is a common feature of random spin chains governed by the IRFP.

This paper is organized as follows. We first propose an improved SDRG procedure, in which a unitary transformation is applied to the operators in each RG step, and show the improved results of the spin correlation function at the ground state. The low-temperature thermodynamic quantities and transport coefficients are evaluated based on the parent Hamiltonian generated by the SDRG transformations in the following two sections. A brief summary and discussions are presented in the last section.

\begin{figure}[tb]
\includegraphics[width=\columnwidth]{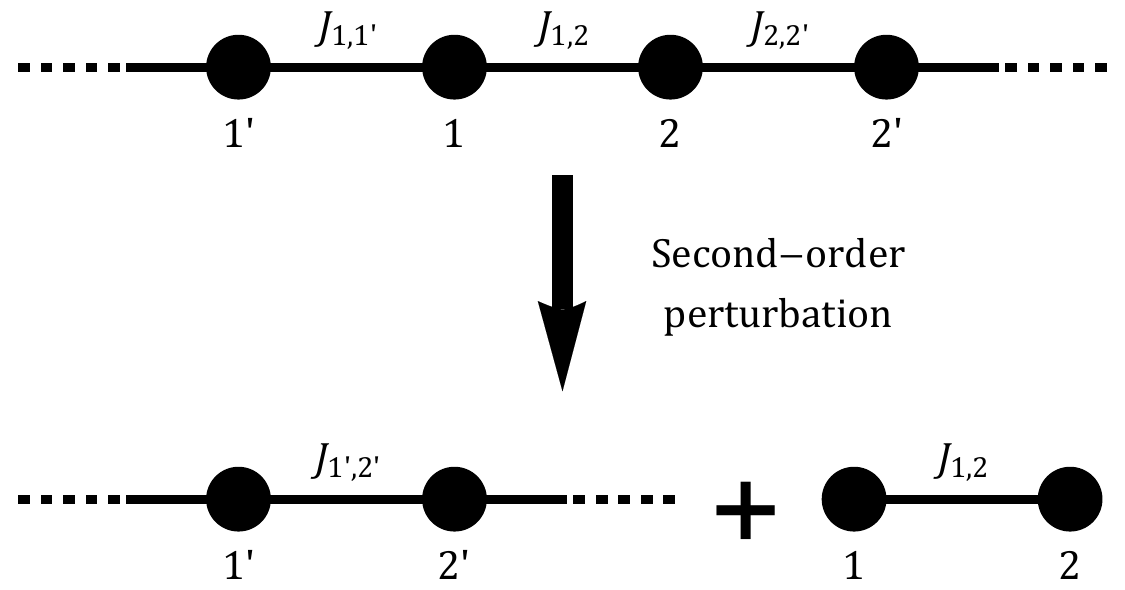}
\caption{Illustration of the real-space SDRG transformation of the random Heisenberg chain. In each RG step, the strongest exchange interaction (labeled by $J_{1,2}$ here) in the remaining spin chain is selected, then the adjacent spins $1$ and $2$ form a singlet state in the low-energy subspace and are decimated from the chain, while the two outer spins $1'$ and $2'$ acquire an effective AF exchange interaction via the second-order perturbation, $J_{1',2'}=J_{1',1}J_{2,2'}/(2J_{1,2})$.}
\label{fig:sdrg}
\end{figure}
	
{\bf Improved SDRG and spin correlation.} The Hamiltonian of the spin-1/2 random Heisenberg chain is given by
\begin{equation}
H=\sum_{l}J_{l,l+1}\mbf{S}_{l}\cdot \mbf{S}_{l+1}\equiv \sum_{l}H_{l,l+1},
\label{eq:ham}
\end{equation}
in which the nearest-neighbor exchange interaction strengths $J_{l,l+1}$'s are independent identically distributed variables. The SDRG procedure is illustrated in Fig. \ref{fig:sdrg}. In each RG step, the largest interaction strength in the spin chain denoted by $J_{1,2}$ is selected, and its two adjacent spins $1$ and $2$ form a singlet state in the low-energy subspace and are decimated from the spin chain. The two outer spins $1'$ and $2'$ acquire an effective AF exchange interaction via the second-order perturbation, $J_{1',2'}=J_{1',1}J_{2,2'}/(2J_{1,2})$. Such RG transformations are iterated for the remaining spin chain until all spins are decimated pairwise and form singlet states in the low-energy subspace. Note that $J_{1',2'}<J_{1,2}$, thus the largest exchange interaction strength decreases monotonically in the RG procedure. The distribution of the logarithm of exchange interaction strengths becomes broader and broader under the RG transformations and approaches the IRFP \cite{Ma1979, Dasgupta1980, Fisher1994}. The ground state is thus approximated by the direct product of these singlet states,
\begin{equation}
\ket{\mr{GS}}=\bigotimes\limits_{\alpha}\frac{1}{\sqrt{2}}\Big(\ket{\uparrow}_{\alpha_{1}}\ket{\downarrow}_{\alpha_{2}}-\ket{\downarrow}_{\alpha_{1}}\ket{\uparrow}_{\alpha_{2}}\Big), \label{eq:rs}
\end{equation}
in which $\alpha$ labels the decimated spin pair $(\alpha_{1}, \alpha_{2})$. Therefore, the ground state is dubbed the random singlet (RS) state \cite{Bhatt1982, Bhatt1992, Fisher1994}. The spin correlation in the RS state comes from the antiparallel alignment within each spin singlet pair. The average inter-sublattice AF correlation decays as $G(r)\propto 1/r^{2}$ at the IRFP \cite{Fisher1994}.

We first introduce an improved SDRG procedure to evaluate the spin correlation function and other physical quantities. In the RG step illustrated in Fig. \ref{fig:sdrg}, while the largest term $H_{1,2}$ splits the Hilbert space into the high-energy triplet and the low-energy singlet subspaces, the subleading terms $H_{1',1}$ and $H_{2,2'}$ contain off-diagonal matrix elements, thus a Schrieffer-Wolff (SW) unitary transformation $U_{1,2}$ is required to eliminate the off-diagonal terms and block-diagonalize the Hamiltonian. The SW transformation up to the $O\big(J_{1',1}/J_{1,2}, J_{2,2'}/J_{1,2}\big)$ order is given by
\begin{equation}
U_{1,2}=e^{-\mbb{P}_{1,2}H_{\mr{off}}/J_{1,2}},
\end{equation}
in which $\mbb{P}_{1,2}=\mbb{T}_{1,2}-\mbb{S}_{1,2}$ with $\mbb{T}_{1,2}$ and $\mbb{S}_{1,2}$ the projection operators into the triplet and the singlet subspaces, respectively; $H_{\mr{off}}=\mbb{T}_{1,2}(H_{1',1}+H_{2,2'})\mbb{S}_{1,2}+\mr{H.c.}$ is the off-diagonal term in the original Hamiltonian. An operator $O$ transforms into
\begin{equation}
U_{1,2}^{-1}OU_{1,2}=O+J_{1,2}^{-1}(\mbb{P}_{1,2}H_{\mr{off}}O-OH_{\mr{off}}\mbb{P}_{1,2}).
\end{equation}
In particular, the Hamiltonian is block-diagonalized and reproduces the second-order perturbation result in the low-energy subspace. The SW transformations are also iterated and only the leading-order nonvanishing terms are retained in each RG step for simplicity.

We then apply this procedure to evaluate the spin correlation function at the ground state, $G_{ij}=\bra{\mr{GS}}S_{i}^{z}S_{j}^{z}\ket{\mr{GS}}$. In the RG step illustrated in Fig. \ref{fig:sdrg}, the operator $S_{i}^{z}S_{j}^{z}$ transforms as follows. First, $S_{1}^{z}S_{2}^{z}$ has a nonvanishing expectation value in the singlet subspace, $G_{12}=-1/4$, thus its subleading correction generated by the SW transformation is neglected. Second, $S_{1}^{z}S_{l}^{z}$ and $S_{2}^{z}S_{l}^{z}$ ($l\neq 1,2$) vanish in the singlet subspace. The SW transformation followed by the projection into the singlet subspace gives
\begin{align}
U_{1,2}^{-1}S_{1}^{z}S_{l}^{z}U_{1,2} &=\frac{1}{2J_{1,2}}\big(-J_{1',1}S_{1'}^{z}+J_{2,2'}S_{2'}^{z}\big)S_{l}^{z},\\
U_{1,2}^{-1}S_{2}^{z}S_{l}^{z}U_{1,2} &=\frac{1}{2J_{1,2}}\big(J_{1',1}S_{1'}^{z}-J_{2,2'}S_{2'}^{z}\big)S_{l}^{z},
\end{align}
thus the correlation function can be obtained from the correlation $\tilde{G}_{ij}$ in the decimated spin chain in the transformed basis,
\begin{align}
G_{1l} &=\frac{1}{2J_{1,2}}\big(-J_{1',1}\tilde{G}_{1',l}+J_{2,2'}\tilde{G}_{2',l}\big),
\label{eq:c1l} \\
G_{2l} &=\frac{1}{2J_{1,2}}\big(J_{1',1}\tilde{G}_{1',l}-J_{2,2'}\tilde{G}_{2',l}\big).
\label{eq:c2l}
\end{align}
Moreover, $S_{l}^{z}S_{m}^{z}$ ($l,m\neq 1,2$) is not affected by the SW transformation, thus
\begin{equation}
G_{lm}=\tilde{G}_{lm}.
\label{eq:clm}
\end{equation}
This procedure is iterated until all spins are decimated and form singlet pairs. Iteratively applying the relations Eqs. (\ref{eq:c1l})--(\ref{eq:clm}) yields an improved estimate of the spin correlation function at the ground state.

\begin{figure}[tb]
\includegraphics[width=\columnwidth]{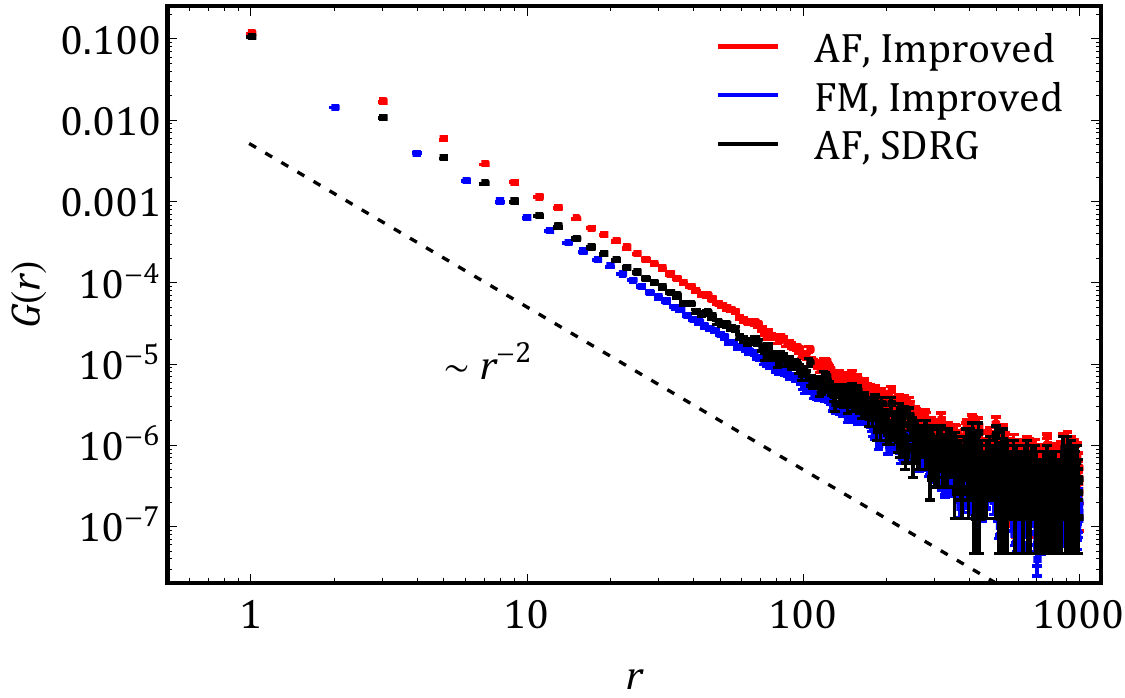}
\caption{The average AF (red symbols) and FM (blue symbols) correlation functions obtained with the improved SDRG procedure, and comparison with the AF correlation function from the conventional SDRG procedure (black symbols). The error bars reflect the statistical uncertainty in the disorder average. The dashed line is guide to the eye.}
\label{fig_G}
\end{figure}

The average spin correlation function at the ground state is numerically calculated for random spin chains with periodic boundary condition. Each spin chain contains $L=2000$ sites, and the results are averaged over $N_{s}=1000$ samples of disorder realization. The long-range and low-energy properties of the random Heisenberg chains are not sensitive to the initial distribution of the exchange interaction strengths, because the distribution always approaches the universal IRFP in the SDRG flow \cite{Fisher1994}. Therefore, we take the following broad probability distribution function $P(J)$ so that the universal scaling behavior of the IRFP can quickly emerge in finite chains,
\begin{equation}
P(J)=
\begin{cases}
1/(J\ln(1/\epsilon)),& \epsilon\leq J\leq 1,\\
0,& \mr{otherwise}.
\end{cases}
\end{equation}
Here, $\epsilon\ll 1/L$ is a small energy scale cutoff. The results of the average spin correlation function are shown and compared with the conventional SDRG result in Fig. \ref{fig_G}. The intra-sublattice FM correlation (blue symbols) is obtained from the iterative transformations in Eqs. (\ref{eq:c1l}) and (\ref{eq:c2l}). Both the intra- and inter-sublattice correlations decay as $1/r^{2}$ in the long-range limit, while the inter-sublattice AF correlation (red symbols) is slightly stronger than the conventional SDRG result (black symbols).

{\bf Low-temperature thermodynamics.} We shall first present an effective description for the low-energy properties of the random spin chain. While the ground state is captured by the RS state in Eq. (\ref{eq:rs}), the low-energy excited states can be obtained by breaking a spin singlet pair into a triplet state, and its excitation energy is the renormalized exchange interaction of the spin pair. In other words, we introduce the following effective Hamiltonian of decoupled spin pairs for the random spin chain,
\begin{equation}
H_{\mr{eff}}=\sum_{\alpha}\tilde{J}_{\alpha}\mbf{S}_{\alpha_{1}}\cdot\mbf{S}_{\alpha_{2}},
\label{eq:heff}
\end{equation}
in which $\alpha$ labels the spin pair $(\alpha_{1}, \alpha_{2})$, and $\tilde{J}_{\alpha}$ denotes the renormalized exchange interaction. This is similar to the (weakly coupled) two-level systems picture of the random transverse-field Ising chain at the IRFP \cite{Fisher1992, Fisher1995, Motrunich2000a, Parameswaran2020}. Each triplet state is actually weakly coupled to its neighboring spins. In the RG step shown in Fig. \ref{fig:sdrg}, the leading-order remnant coupling is given by
\begin{equation}
\mbb{T}_{1,2}(H_{1',1}+H_{2,2'})\mbb{T}_{1,2}=\frac{1}{2}\big(J_{1',1}\mbf{S}_{1'}+J_{2,2'}\mbf{S}_{2'}\big)\cdot\mbf{S}_{1,2},
\end{equation}
in which $\mbf{S}_{1,2}$ is the total spin operator of the pair $(1,2)$ in its triplet subspace. The coupling is weaker than the triplet excitation gap $J_{1,2}$ by a factor $O(J_{1',1}/J_{1,2},J_{2,2'}/J_{1,2})$, which is vanishingly small approaching the IRFP in the low-energy limit. Therefore, the parent Hamiltonian Eq. (\ref{eq:heff}) should be asymptotically effective for the low-temperature properties of the random Heisenberg chain.

\begin{figure}[tb]
\includegraphics[width=\columnwidth]{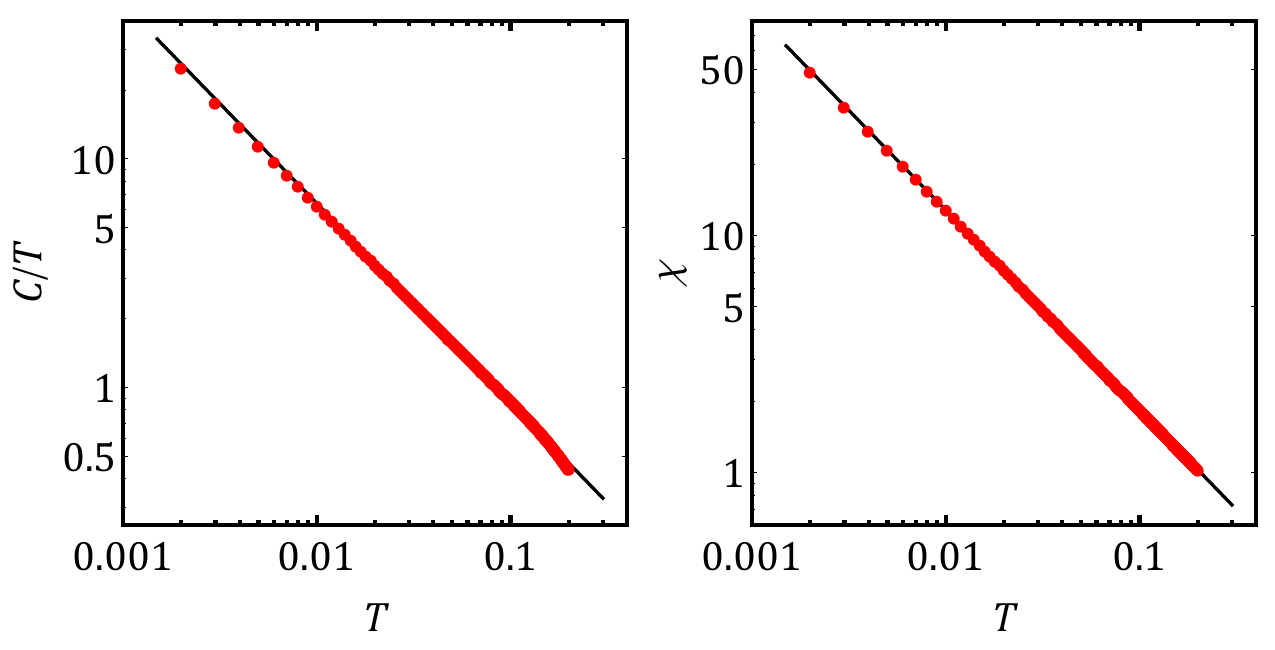}
\caption{Low-temperature thermodynamic quantities: Specific heat coefficient $C/T$ (left panel) and uniform magnetic susceptibility $\chi$ (right panel). The power-law fitting yields $C/T\propto T^{-0.873(2)}$ and $\chi\propto T^{-0.841(1)}$.}
\label{fig:thermal}
\end{figure}
	
We then adopt the effective Hamiltonian to calculate the low-temperature thermodynamic properties, which will be compared with the transport coefficients later. The partition function of the decoupled spin pairs at temperature $T=1/\beta$ is
\begin{equation}
Z=\prod_{\alpha}\big(1+3e^{-\beta\tilde{J}_{\alpha}}\big).
\end{equation}
The specific heat is given by
\begin{equation}
C=\frac{3\beta^{2}}{L}\sum_{\alpha}\frac{\tilde{J}_{\alpha}^{2}e^{\beta\tilde{J}_{\alpha}}}{(3+e^{\beta\tilde{J}_{\alpha}})^{2}},
\end{equation}
and the uniform magnetic susceptibility,
\begin{equation}
\chi=\frac{2\beta}{L}\sum_{\alpha}\frac{1}{3+e^{\beta\tilde{J}_{\alpha}}}.
\end{equation}
The numerical results are presented in Fig. \ref{fig:thermal}. Both quantities show the power-law scaling at low temperature, $C/T\propto T^{-0.873(2)}$ and $\chi\propto T^{-0.841(1)}$. This scaling form is consistent with previous SDRG results \cite{Ma1979, Dasgupta1980, Hirsch1980, Hirsch1980a, Bhatt1981a, Bhatt1982} and indicates a divergent low-energy density of states in the random spin chain.

{\bf Low-temperature transport.} We then apply the effective Hamiltonian to the low-temperature transport coefficients, which can be calculated from the correlation functions of conserved current operators. For a conserved symmetry charge density $\rho_{l}$, the current operator $j_{l}$ is constructed with the continuity equation, $\dot{\rho}_{l}=i[H,\rho_{l}]=-(j_{l}-j_{l-1})$. For the random Heisenberg chain Eq. (\ref{eq:ham}), the spin current operator is
\begin{equation}
j_{l}^{s}=\frac{1}{2i}J_{l,l+1}\big(S_{l}^{-}S_{l+1}^{+}-S_{l}^{+}S_{l+1}^{-}\big),
\end{equation}
and the energy current operator is
\begin{equation}
j_{l}^{E}=J_{l,l+1}J_{l+1,l+2}\mbf{S}_{l}\cdot(\mbf{S}_{l+1}\times\mbf{S}_{l+2}).
\end{equation}
The Kubo formula of the dynamical spin conductivity is \cite{Kubo1966, Mahan2000many}
\begin{equation}
\sigma_{s}(\omega)=\frac{1}{L}\sum_{lm}\int_{0}^{\infty}\dd t\, e^{i\omega t} \int_{0}^{\beta}\dd\lambda\,\langle j_{l}^{s}(-i\lambda)j_{m}^{s}(t)\rangle,
\end{equation}
in which $\langle\cdots\rangle$ is the thermal average. The dc spin conductivity can be cast into the Lehmann's spectral representation,
\begin{equation}
\sigma_{s}=\frac{\pi\beta}{ZL}\sum_{lm}\sum\limits_{\mu\nu}^{E_\mu=E_\nu}e^{-\beta E_\mu} \langle\mu|j_{l}^{s}|\nu\rangle\langle\nu|j_{m}^{s}|\mu\rangle,
\label{eq:lehmann1}
\end{equation}
in which $Z$ is the partition function. The second summation is taken over degenerate eigenstates $\ket{\mu}$ and $\ket{\nu}$. The thermal conductivity can be derived by treating the spatial modulation of temperature $\delta T(r)$ as an external perturbation, $V=-\beta\int \dd r\, h(r)\delta T(r)$, and cast into the Lehmann's representation,
\begin{equation}
\kappa=\frac{\pi\beta^2}{ZL}\sum_{lm}\sum_{\mu\nu}^{E_\mu=E_\nu}e^{-\beta E_\mu} \langle\mu|j_{l}^{E}|\nu\rangle\langle\nu|j_{m}^{E}|\mu\rangle.
\label{eq:lehmann2}
\end{equation}

\begin{figure}[tb]
\includegraphics[width=\columnwidth]{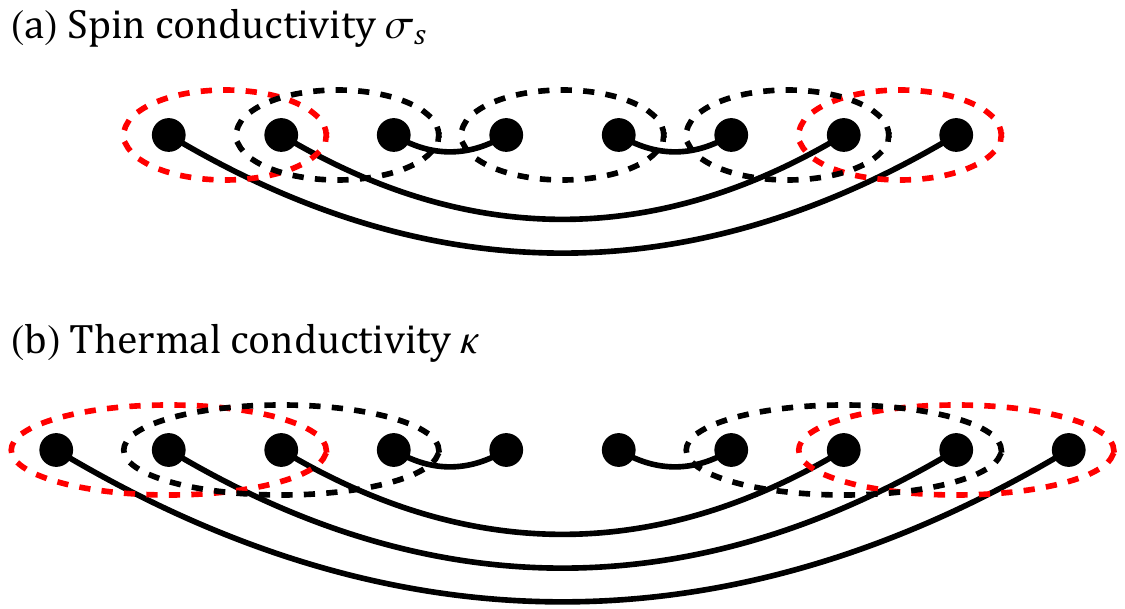}
\caption{Illustration of nonzero contributions to the spin and the thermal conductivities. Paired spins in the effective Hamiltonian are joined by thick arcs. For the spin conductivity (upper panel), unpaired nearest-neighbor spins encircled by dashed red and black ellipses contribute to the first positive term in Eq. (\ref{eq:sigma}), while the two red ellipses highlight a negative contribution to the second term. Similar labels are used for the thermal conductivity in Eq. (\ref{eq:kappa}) (lower panel).}
\label{fig:kubo}
\end{figure}

In the effective Hamiltonian Eq. (\ref{eq:heff}), the eigenstates are the direct product of singlet or triplet spin pairs. If we neglect the accidental degeneracy, the degenerate states appearing in Eqs. (\ref{eq:lehmann1}) and (\ref{eq:lehmann2}) come from the spin triplet subspace of the spin pairs, thus only the block-diagonal elements of the current operators are needed,
\begin{equation}
j_{\mr{eff}}=\mbb{T}_{1,2}j\mbb{T}_{1,2}+\mbb{S}_{1,2}j\mbb{S}_{1,2}.
\end{equation}
Moreover, the SW transformation in the RG step does not alter the spin and the energy current operators up to the leading order, $\big(U_{1,2}^{-1}j_{l}^{s/E}U_{1,2}\big)_{\mr{eff}}=j_{l}^{s/E}$. Using the direct-product form of the effective Hamiltonian eigenstates, the spin conductivity is given by
\begin{equation}
\sigma_{s}=\frac{\pi\beta}{2L}\Big(\sum_{l}{}^{\prime} J_{l,l+1}^2 f_{l}f_{l+1}-\sum_{lm}{}^{\prime} J_{l,l+1}J_{m,m+1}f_{l}f_{m}\Big),
\label{eq:sigma}
\end{equation}
in which $f_{l}=(3+e^{\beta\tilde{J}_{l}})^{-1}$, and $\tilde{J}_{l}$ is the renormalized exchange interaction of site $l$ and its partner in the effective Hamiltonian. Here, the nonzero contributing terms are illustrated in Fig. \ref{fig:kubo} (a), in which the summation $\sum_{l}^\prime$ is taken over site indices $l$ such that the spins $\mbf{S}_{l}$ and $\mbf{S}_{l+1}$ are not paired up in the effective Hamiltonian, and $\sum_{lm}^\prime$ is over indices $l\neq m$ such that $(l,m+1)$ and $(l+1,m)$ are paired up, respectively. The thermal conductivity is given by
\begin{equation}
\begin{split}
\kappa&=\frac{3\pi\beta^{2}}{4L}\Big(\sum_l{}^{''}J_{l,l+1}^{2} J_{l+1,l+2}^{2} f_{l} f_{l+1} f_{l+2} \\ 
&-\sum_{lm}{}^{''}J_{l,l+1} J_{l+1,l+2}J_{m,m+1} J_{m+1,m+2} f_{l} f_{l+1}f_{l+2}\Big).
\label{eq:kappa}
\end{split}
\end{equation}
Here, as illustrated in Fig. \ref{fig:kubo} (b), the summation $\sum_{l}^{''}$ is taken over indices $l$ such that no spin pair is formed within $\mbf{S}_{l}$, $\mbf{S}_{l+1}$ and $\mbf{S}_{l+2}$, while $\sum_{lm}^{''}$ is over indices $l\neq m$ such that $(l+2,m)$, $(l+1,m+1)$ and $(l,m+2)$ are paired up, respectively.

\begin{figure}[tb]
\includegraphics[width=\columnwidth]{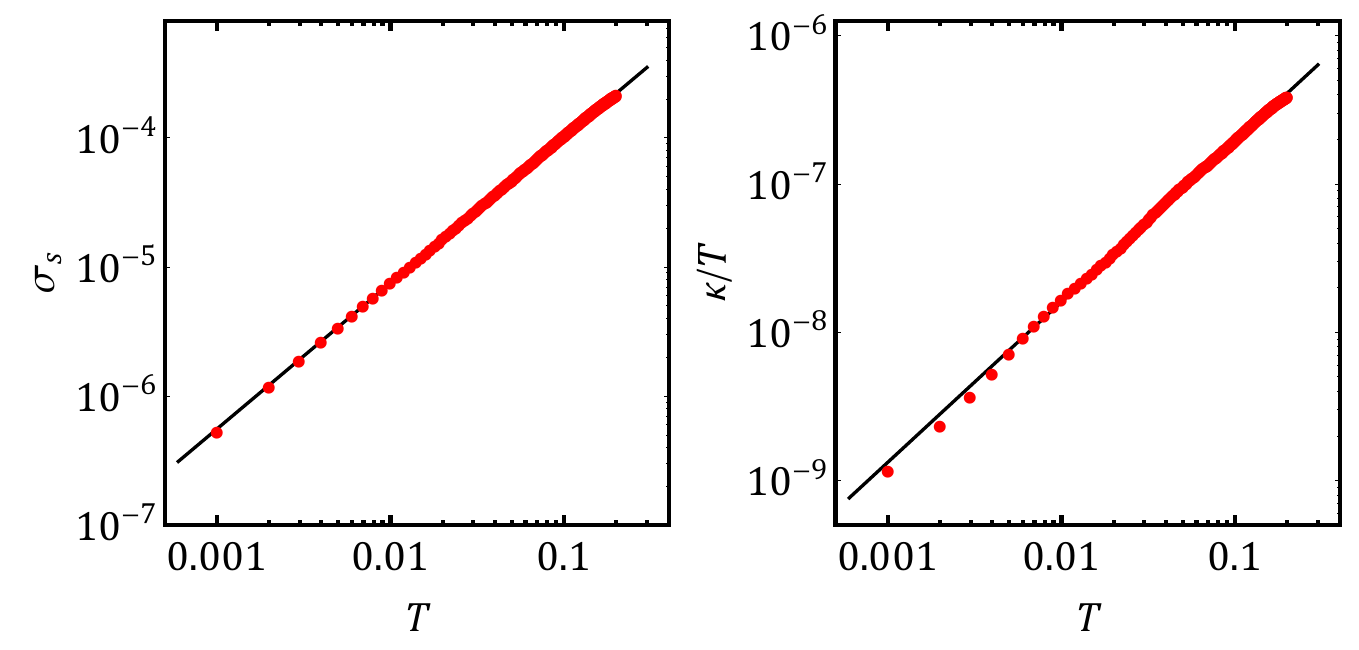}
\caption{Low-temperature transport coefficients: Spin conductivity $\sigma_{s}$ (left panel) and thermal conductivity coefficient $\kappa/T$ (right panel). The power-law fitting yields $\sigma_{s}\propto T^{1.130(1)}$ and $\kappa/T\propto T^{1.082(3)}$.}
\label{fig:trans}
\end{figure}

Numerical results of the low-temperature spin and thermal conductivities are plotted in Fig. \ref{fig:trans}. Both show power-law scaling with the temperature, $\sigma_{s}\propto T^{1.130(1)}$ and $\kappa/T\propto T^{1.082(3)}$. This is in sharp contrast to the divergence of $C/T$ and $\chi$ in the low-temperature limit, and implies that despite the divergent low-energy density of states, most of these states are localized in the low-energy limit. The vanishing of transport coefficients is consistent with the low-temperature thermal transport experiment in the quasi-1D antiferromagnet copper benzoate.

{\bf Summary and discussions.} To summarize, we adopt an effective Hamiltonian consisting of decoupled spin pairs generated in the SDRG procedure to study the low-temperature properties of spin-1/2 random Heisenberg chains. We find a sharp disparity between the thermodynamic and transport coefficients. Despite the divergent low-energy density of states indicated by the specific heat coefficient $C/T\sim T^{-\alpha_{c}}$ and the magnetic susceptibility $\chi\sim T^{-\alpha_{s}}$ with $0<\alpha_{c,s}<1$, the spin and the thermal conductivity follow the scaling $\sigma_{s}\sim T$ and $\kappa/T\sim T$, which vanish in the low-temperature limit and imply that extended states are rare in the low-energy limit.

The random transverse-field Ising model (TFIM) at its critical point and the random spin-$1/2$ XY model in one dimension also flow to the IRFP under the SDRG transformations with divergent low-energy density of states \cite{Igloi2005}. These models map to free fermions with random hopping (and pairing in the random TFIM) and random onsite chemical potential by the Jordan-Wigner transformation, thus their eigenstates are also localized according to the scaling theory of Anderson localization \cite{Abrahams1979} and cannot conduct heat or spin in the low-temperature limit. Therefore, there is also a disparity between the vanishing transport coefficients and the divergent low-energy density of states. We thus conjecture that such a disparity is a common feature of random spin chains controlled by the IRFP in the low-energy limit.

While the effective Hamiltonian captures the RS ground state, its validity for the excited states deserves more attention. Even though the remnant coupling of the spin triplet pair with its neighboring spins is weaker than the excitation energy by a vanishingly small factor in the low-energy limit, the coupling cannot be neglected and could finally delocalize the triplet excitations at higher temperature. Moreover, weakly random spin chains must show similar properties as the corresponding uniform system at temperature higher than the disorder strength. The crossover between the low- and the high-temperature regimes is an interesting issue. We may either treat the coupling of excited states as a perturbation to the effective Hamiltonian, or study certain exactly solvable models to examine the validity of the effective Hamiltonian approach and delineate the crossover between the low- and high-temperature behavior in future works. With results in both low- and high-temperature regimes, we may gain a complete understanding of the transport properties of disordered quantum spin chains.

\acknowledgments
This work is supported by the National Natural Science Foundation of China (Grant No. 12174387), Chinese Academy of Sciences (Nos. YSBR-057 and JZHKYPT-2021-08), and the Innovative Program for Quantum Science and Technology (No. 2021ZD0302600).

\bibliography{../../BibTex/library,../../BibTex/books}
\end{document}